\documentclass[11pt, preprint]{article}
\usepackage{color} \usepackage{cancel}
\usepackage{float} \usepackage{graphicx}
\usepackage{subfigure} \usepackage{caption}	
\usepackage{dcolumn}
\usepackage[toc,page]{appendix} \usepackage{authblk}	 		
\usepackage{indentfirst}	
\usepackage{authblk}	 		
\usepackage{multirow}     
\usepackage{hyperref}     
\usepackage{amsfonts, amssymb, amsmath}
\usepackage[left=2.5cm,right=2.5cm,top=2.9cm,bottom=2.9cm,a4paper]{geometry}
\usepackage[noadjust]{cite}
\usepackage{filecontents}

\usepackage{soul}
\usepackage{xcolor}
\setstcolor{red}

\graphicspath{{./figures/}}
\hypersetup{ colorlinks   = true, 
urlcolor     = blue, 
linkcolor    = blue, 
citecolor    = red 	 
}

\title{\Large\bf Reconstruction of  GR cosmological solutions in modified gravity theories}
\author[1]{I. V. Fomin\thanks{ingvor@inbox.ru}}
\author[1,2]{S. V. Chervon\thanks{chervon.sergey@gmail.com}}
\affil[1]{\small \it  Bauman Moscow State Technical University, 2-nd Baumanskaya street, 5, Moscow, 105005, Russia}
\affil[2]{\small \it Kazan Federal University, Kremlevskaya street 18, Kazan, 420008, Russia}

\begin{document}

\maketitle
\begin{abstract}
We study the special class of the exact solutions in cosmological models based on the Generalized Scalar-Tensor Gravity with non-minimal coupling of a scalar field to the Ricci scalar and to the Gauss-Bonnet scalar in 4D Friedmann universe corresponding to similar ones in GR. The parameters of cosmological perturbations in such models correspond to the case of Einstein gravity with a high precision. As the example of proposed approach, we obtain the exact solutions for the power-law and exponential power-law inflation.
\end{abstract}
\section{Introduction}

An inflationary scenario was invented to solve many problems in big-bang cosmology,
including the explanation for the origin of large scale structure of the universe.
The first models of cosmological inflation as a rule\footnote{We have to stress that Starobinsky model \cite{Starobinsky:1980te} was related to modified f(R) gravity.} were built on the basis of General Relativity (GR) in 4D Friedmann-Robertson-Walker (FRW) space-time under the assumption of the existence of self-interacting scalar field which is the source of the accelerated expansion of the universe~\cite{Starobinsky:1980te,Guth:1980zm,Linde:1981mu,Linde:2005ht,Lyth:2009zz}.

After the discovery of the second accelerated expansion of the universe at the present time, cosmological models with various modifications of Einstein gravity were proposed for its explanation or, in other words, to explain the nature of dark energy which include, among others, scalar-tensor gravity theories and Einstein-Gauss-Bonnet gravity~\cite{Frieman:2008sn,Nojiri:2007uq,Clifton:2011jh}.

Nevertheless, recent observations of gravitational waves from the merger of neutron stars lead to serious restrictions on the modification of Einstein's theory of gravity. These restrictions are related to the velocity of the propagation of gravitational waves
which is equal to speed of light in vacuum and which corresponds to the case of GR  with the high accuracy $10^{-15}$~\cite{Monitor:2017mdv}.

Scalar-tensor gravity (STG) with non-minimal coupling of a scalar field to curvature are important extensions of GR explaining the initial inflationary evolution, as well as the late accelerating expansion of the universe~\cite{Fujii,Faraoni}.
The standard method for analysis of STG models and studying correspondence to GR is conformal transformations from Jordan frame to Einstein one~\cite{Fujii,Faraoni}. The connection between STG and GR cosmological solutions without conformal transformations was considered in~\cite{Fomin:2018blx}. Also, the influence of non-minimal coupling on the deviation from de Sitter expansion is studied in~\cite{Fomin:2017sbt} and it was shown that this approach leads to the models which have a good concordance to the observational constraints.


For the very early universe approaching the Planck scale one can consider Einstein gravity with some corrections as the effective theory of the quantum gravity. The effective supergravity action from superstrings induces correction terms of higher order in the curvature, which may play a significant role in the early Universe. The one of such correction is the Gauss-Bonnet (GB) term in the low-energy effective action of the heterotic strings~\cite{Zwiebach:1985uq}.
Also, the GB term arises in the second order of the Lovelock gravity which is the generalization of the Einstein gravity~\cite{Lovelock:1971yv}.

The exact solutions for cosmological models on the basis of Einstein-Gauss-Bonnet gravity with non-minimal coupling of a scalar field with GB-scalar in 4D Friedmann universe were considered in~\cite{Nojiri:2006je,Calcagni:2005im,Calcagni:2006ye,Cognola:2006sp,
Nojiri:2005vv,Nojiri:2005jg,Nojiri:2007te,Bamba:2014zoa,Guo:2009uk,
Guo:2010jr,Jiang:2013gza,Koh:2014bka,Koh:2016abf,Kanti:2015pda,vandeBruck:2017voa,
Kanti:1998jd,Nozari:2017rta,Chakraborty:2018scm,Odintsov:2018zhw}.
%
When such cosmological models are considering the
two important problems can be noted: the lack of conformal transformations to Einstein frame and the dependence of the propagation velocity of gravitational waves on time.

To avoid the first problem, the connection between cosmological models based on standard GR inflation
and EGB inflation was found in papers~\cite{Fomin:2017vae,Fomin:2017qta,Fomin:2018typ} without
conformal transformations. Therefore, one has the possibility to compare these
two types of models and evaluate the impact of non-minimal interaction on the character of cosmological inflation.

To solve the second problem we suggest, in the present paper, the connection between cosmological dynamics of EGB and standard GR inflation which allows rapid transform to general relativity case during inflationary epoch.
Based on suggested connection it is possible to construct EGB inflationary models with fast approaching to GR inflation.

To cover more types of modified gravity models considering in literature it is useful to generalize developed approach on models with non-minimal interaction of a scalar field with Ricci and Gauss-Bonnet scalars simultaneously, which can be considered both together and separately. We will call such models as Generalized Scalar-Tensor Gravity (GSTG) theories.

Since the most common scalar-tensor theory of gravity is the Hordeski gravity~\cite{Horndeski:1974wa,Starobinsky:2016kua,Capozziello:2018gms,Mizuno:2010ag,DeFelice:2011uc}, it is also necessary to determine the relationship between the types of gravity under consideration and GSTG theory. Such a connection was presented in the works~\cite{Mizuno:2010ag,DeFelice:2011uc}.

One more issue we discuss in the present article is an appearance of various normalizations of the tensor of gravitational waves in the literature. We discuss the influence of these normalizations on the values of the parameters of cosmological perturbations and the verification of inflationary models.


Our paper is organized as follows. In Section 2 we represent the equations of cosmological dynamics for GSTG in spatially-flat 4D Friedmann universe and we introduce the deviation parameters which characterize the departure of non-minimal coupling functions from zero (i.e. the absence of non-minimal interactions). It is shown there the way of obtaining GSTG model parameters which reconstruct any exact solution from GR cosmology.  With suggested connections between deviation parameters and the scale factor of the universe expansion it is proved fast approaching GSTG inflation to GR one with high accuracy.
The Section 3 is devoted to consideration of GSTG cosmological parameters of cosmological perturbations and comparison of them to observational constraints. Also possible normalizations of tensor perturbations are discussed as well there. In Section 4 we apply proposed approach to generate new exact inflationary solutions for GSTG from known solutions in scalar GR cosmology: the power-law and exponential power-law scale factor.
The corresponding of obtained GSTG power-law solution to observational constraints is performed as well. In Section 5 we study the connection of GSTG to Horndeski gravity and find the relation between Horndeski gravity parameters and ones for GSTG.
In Section 6 we briefly discuss the restrictions on the deviations from GR for gravity models under consideration.
Finally, Section 7 contains our summary.

\section{The conformity between GR and GSTG motivated inflation}

We study GSTG theory with the action
\begin{equation}
\label{actionh}
S=\frac{1}{2}\int d^4x\sqrt{-g}\left[F(\phi,R) - \omega(\phi)g^{\mu\nu}\partial_{\mu}\phi \partial_{\nu}\phi-2V(\phi)\right],
\end{equation}
where
\begin{equation}
\label{FRF}
F(\phi,R)=R+f(\phi)R+\xi(\phi)R^{2}_{GB}
\end{equation}
Here $f(\phi)$ defines a non-minimal coupling of a scalar field with Ricci scalar and $\xi(\phi)$ defines a non-minimal coupling of a scalar field with Gauss-Bonnet scalar $R^{2}_{GB} = R_{\mu\nu\rho\sigma} R^{\mu\nu\rho\sigma}- 4 R_{\mu\nu} R^{\mu\nu} + R^2.$

The equations of cosmological dynamics at the stage of inflation in spatially flat 4D Friedmann universe (in the system of units $m^{2}_{P}=c=1$)
\begin{equation}
ds^2=-dt^2+a^{2}(t)\left(dx^{2}+dy^{2}+dz^{2}\right),
\end{equation}
can be written as
\begin{eqnarray} \label{beq2ah}
&&E_{1}\equiv3(1+f)H^{2}+3H\dot{f}-\frac{\omega}{2}\dot{\phi}^{2}-V(\phi)-12H^{3}\dot{\xi}=0,\\
\label{beq3ah}
&&E_{2}\equiv(1+f)(3H^{2}+2\dot{H})+2H\dot{f}+\ddot{f}+\frac{\omega}{2}\dot{\phi}^{2}-
V(\phi)
-8H^{3}\dot{\xi}-8H\dot{H}\dot{\xi}-4H^{2}\ddot{\xi}=0,\\
&&E_{3}\equiv\omega\ddot{\phi} + 3\omega H\dot{\phi} + \frac{1}{2}\dot{\phi}^{2}\omega'_{\phi}+V'_{\phi}-6H^{2}f'_{\phi}-3\dot{H}F'_{\phi}
+12H^{4}\xi'_{\phi}+12H^{2}\dot{H}\xi'_{\phi}= 0, \label{beq4ah}
\end{eqnarray}
with the additional condition~\cite{DeFelice:2011jm}
\begin{equation} \label{con123h}
\dot{\phi}E_{3}+\dot{E}_{1}+3H(E_{1}-E_{2})=0.
\end{equation}
Taking into account \eqref{con123h} we conclude that two equations from (\ref{beq2ah})--(\ref{beq4ah}) are independent only.

Our choice is to represent the cosmological dynamic equations as
\begin{equation}
 \label{EQG1}
V(\phi)=3(1+f)H^{2}+(1+f)\dot{H}+\frac{5}{2}H\dot{f}+\frac{1}{2}\ddot{f}
-10H^{3}\dot{\xi}-2H^{2}\ddot{\xi}-4H\dot{H}\dot{\xi},
\end{equation}
\begin{equation}
\label{EQG2}
\omega(\phi)\dot{\phi}^{2}=H\dot{f}-2(1+f)\dot{H}-\ddot{f}-4H^{3}\dot{\xi}+8H\dot{H}\dot{\xi}+4H^{2}\ddot{\xi}.
\end{equation}

As one can see, the constant coupling of the scalar field and the Gauss-Bonnet scalar $\xi=const$ doesn't change the equations of cosmological dynamics for scalar-tensor gravity.

The case of GR corresponds to the choice $f=0$, $\omega=1$ and $\xi=0$ . Then from the (\ref{actionh}) one derives the action
\begin{equation}
\label{actionE}
S_{E}=\int d^{4}x\sqrt{-g}\left[\frac{1}{2}R-\frac{1}{2}g^{\mu\nu}\partial_{\mu}\phi\partial_{\nu}\phi -V(\phi)\right],
\end{equation}
which leads to the following cosmological dynamic equations
\begin{eqnarray}
 \label{DE1}
&&3H^{2}=\frac{1}{2}\dot{\phi}^{2}+V(\phi),\\
\label{DE2}
&&-3H^{2}-2\dot{H}=\frac{1}{2}\dot{\phi}^{2}-V(\phi),\\
\label{DE3}
 &&\ddot{\phi} + 3H\dot{\phi} +V'_{\phi}= 0,
\end{eqnarray}

We can rewrite them in terms of the cosmic time and the scalar field as arguments, in the form
\begin{eqnarray}
\label{beq1}
&&V_{E}(\phi(t))=3H^{2}+\dot{H},~~~~V_{E}(\phi)=3H^{2}-2H'^{2}_{\phi},\\
\label{beq2}
&&\dot{\phi}^{2}_{E}=-2\dot{H},~~~~~~~~~~~~~~~~~~\dot{\phi}_{E}=-2H'_{\phi}.
\end{eqnarray}

To characterize the difference between cosmological inflation based on Einstein gravity and Generalized Scalar-Tensor Gravity, we introduce the {\it deviation parameters} $\Delta_{ST}=\Delta_{ST}(t)$  and $\Delta_{GB}=\Delta_{GB}(t)$ which are connected to the coupling functions as
\begin{eqnarray}
\label{fST}
f(\phi)=-\Delta_{ST},\\
\label{xiGB}
\dot{\xi}=-\frac{\Delta_{GB}}{2H^{2}}.
\end{eqnarray}

In terms of these parameters one has the following background dynamic equations
\begin{eqnarray}
\label{BSEM1GEN}
&&V(\phi)=3(1-\Delta_{ST})H^{2}+(1-\Delta_{ST})\dot{H}-\frac{5}{2}H\dot{\Delta}_{ST}
-\frac{1}{2}\ddot{\Delta}_{ST}+\dot{\Delta}_{GB}+5H\Delta_{GB},\\
\label{BSEM2GEN}
&&\omega(\phi)\dot{\phi}^{2}=-2(1-\Delta_{ST})\dot{H}-H\dot{\Delta}_{ST}+\ddot{\Delta}_{ST}
-2\dot{\Delta}_{GB}+2H\Delta_{GB}.
\end{eqnarray}

For the case $\Delta_{ST}=0$  and $\Delta_{GB}=0$ the equations (\ref{BSEM1GEN}) and (\ref{BSEM2GEN}) are transformed to (\ref{beq1})--(\ref{beq2}).

To analyze the variety  of exact solutions of cosmological inflation models based on the GSTG with the dynamic equations  (\ref{EQG1})--(\ref{EQG2}) or (\ref{BSEM1GEN})--(\ref{BSEM2GEN}) we prove the following assertion
on a special class of the exact cosmological solutions:

 {\it In four dimensional Friedman-Robertson-Walker space-time for each exact background cosmological solution in GR cosmology, where the scalar field $\phi\neq const,$ exists the same solution in the GSTG with nontrivial in general case coupling functions $f\neq0$, $\xi\neq const$ and kinetic function $\omega\neq1$.}

To prove this assertion we define the kinetic function and the deviation parameters as follows
\begin{equation}
\label{omegaSTGB}
\omega(\phi)=1+\frac{3}{\epsilon}\left(\Delta_{ST}+2\frac{\Delta_{GB}}{H}\right),
\end{equation}
\begin{equation}
\label{DEVST}
\Delta_{ST}(t)=\beta_{ST}a^{-2}(t),
\end{equation}
\begin{equation}
\label{DEVGB}
\Delta_{GB}(t)=\alpha_{GB}a^{-5}(t),
\end{equation}
where $\epsilon=-\dot{H}/H^{2}$ is the slow-roll parameter, $\beta_{ST}$ and $\alpha_{GB}$ are the coupling constants of the scalar field with the Ricci scalar and Gauss-Bonnet scalar, respectively.

 It is not difficult to check that after substituting the functions (\ref{omegaSTGB})--(\ref{DEVGB}) into the background dynamics equations (\ref{BSEM1GEN})--(\ref{BSEM2GEN}) we have the same equations as (\ref{beq1})--(\ref{beq2})
\begin{eqnarray}
\label{E1GRLIKE}
&&V(\phi)=3H^{2}+\dot{H},\\
\label{E2GRLIKE}
&&\dot{\phi}^{2}=-2\dot{H}.
\end{eqnarray}

Further, from \eqref{fST} and \eqref{xiGB} we derive the expressions for non-minimal coupling functions and the kinetic function
\begin{eqnarray}
\label{PAR1}
&&f(t)=-\frac{\beta_{ST}}{a^{2}(t)},\\
\label{PAR1-2}
&&\dot{\xi}=-\frac{\alpha_{GB}}{2a^{5}H^{2}},\\
\label{PAR1-3}
&&\omega(t)=1+\frac{3}{\epsilon a^{2}}\left(\beta_{ST}+\frac{2\alpha_{GB}}{Ha^{3}}\right).
\end{eqnarray}

Therefore, for each exact solution of the system (\ref{E1GRLIKE})--(\ref{E2GRLIKE}) one can find corresponding
functions (\ref{PAR1})-(\ref{PAR1-3}) which characterise the type of GSTG model.

Now we want to represent the equations (\ref{E1GRLIKE})--(\ref{E2GRLIKE}) and the functions (\ref{PAR1})-(\ref{PAR1-3}) in terms
of a scalar field $\phi$ as the argument on the basis of the following relations
\begin{eqnarray}
&&\dot{H}=-2H'^{2}_{\phi},\\
&&\dot{\xi}=\xi'_{\phi}\dot{\phi}=-2\xi'_{\phi}H'_{\phi},\\
&&a(\phi)=a_{0}\exp\left(-\frac{1}{2}\int\frac{H}{H'_{\phi}}d\phi\right).
\end{eqnarray}
As the result, equations (\ref{E1GRLIKE})--(\ref{E2GRLIKE}) are transformed to we have the Ivanov-Salopek-Bond equations~\cite{givanov81,Salopek:1990re}
\begin{eqnarray}
\label{E1GRLIKEF}
&&V(\phi)=3H^{2}-2H'^{2}_{\phi},\\
\label{E2GRLIKEF}
&&\dot{\phi}=-2H'_{\phi},
\end{eqnarray}
with corresponding non-minimal coupling and kinetic functions
\begin{eqnarray}
\label{RCOR}
&& f(\phi)=-\frac{\beta_{ST}}{a^{2}_{0}}\exp\left(\int\frac{H}{H'_{\phi}}d\phi\right), \\
&&\xi'_{\phi}=\frac{\alpha_{GB}}{4a^{5}_{0}H'_{\phi}H^{2}}\exp\left(\frac{5}{2}\int\frac{H}{H'_{\phi}}d\phi\right),\\
\label{RCOR1}
&&\omega(\phi)=1+\frac{3}{2}\left(\frac{H}{H'_{\phi}}\right)^{2}\exp\left(\int\frac{H}{H'_{\phi}}d\phi\right)
\left[\beta_{ST}+\frac{2\alpha_{GB}}{H}\exp\left(\frac{3}{2}\int\frac{H}{H'_{\phi}}d\phi\right)\right].
\end{eqnarray}

Also, using the structure of \eqref{RCOR1} one can represent the coupling function $\xi(\phi)$ as
\begin{equation}
\label{Psi1}
\xi(\phi)=\Psi(\phi)\exp\left(\frac{5}{2}\int\frac{H}{H'_{\phi}}d\phi\right),
\end{equation}
where the function $\Psi(\phi)$ is defined from the equation
\begin{equation}
\label{Psi2}
\left(\Psi'_{\phi}H'_{\phi}+\frac{5}{2}\Psi H\right)H^{2}=\frac{\alpha_{GB}}{4a^{5}_{0}}.
\end{equation}

For minimal coupling of a scalar field and the Gauss-Bonnet term $\alpha_{GB}=0$ from the equation (\ref{Psi2}) one has
\begin{equation}
\Psi(\phi)=\text{const}\times\exp\left(-\frac{5}{2}\int\frac{H}{H'_{\phi}}d\phi\right),
\end{equation}
and, therefore, from (\ref{Psi1}) we obtain $\xi=const$.

We will consider the equation (\ref{Psi2}) as an explicit integrability condition of the system (\ref{beq2ah})--(\ref{beq4ah}) for GR-like cosmological models based on GST gravity.

Further, we consider the evolution of the deviation parameters in terms of the $e$-folds numbers $N=\ln\left(a/a_{0}\right)=-(1/2)\int \left( H/H' \right)d\phi $.
From the expressions (\ref{DEVST})--(\ref{DEVGB}) we obtain $\Delta_{ST}\propto\exp(-2N)$ and $\Delta_{GB}\propto\exp(-5N)$. Thus, the initial deviations between GR and GST gravity rapidly decrease with the expansion of the universe in these models.
The value of  the $e$-folds numbers at the end of inflation is estimated as $N=50-60$ and, therefore, we have
\begin{eqnarray}
\label{dev}
\frac{\Delta_{ST}(N=60)}{\Delta_{ST}(N=0)}=e^{-120}\approx7.7\times10^{-53},~~~~
\frac{\Delta_{GB}(N=60)}{\Delta_{GB}(N=0)}=e^{-300}\approx5.2\times10^{-131},
\end{eqnarray}
where $\Delta_{ST}(N=0)$, $\Delta_{GB}(N=0)$ and $\Delta_{GB}(N=60)$, $\Delta_{GB}(N=60)$ correspond to the values
of the deviation parameters at the beginning and at the end of inflation. Therefore, for the case $\Delta_{ST}(N=0)\ll e^{120}$
(for example, for conformal coupling with $(\beta_{ST}/a^{2}_{0})=1/6)$ and $\Delta_{GB}(N=0)\ll e^{300}$ at the end of inflation one has $\Delta_{ST}\simeq0$ and $\Delta_{GB}\simeq0$.

\section{The parameters of cosmological perturbations}

One of the main methods of verification of cosmological models is the comparison of the obtained parameters of cosmological perturbations with observational constraints.
Now, we consider the parameters of cosmological perturbations in the inflationary models based on modified gravity~\cite{DeFelice:2011uc,DeFelice:2011jm} in terms of the deviation parameters for decreasing slow-roll parameters such that at the crossing of the Hubble radius one has the conditions $\epsilon\ll 1$ and $\delta\ll 1$ (inflation occurs under the condition $0\leq\epsilon<1$ and ends when $\epsilon=1$).

To calculate the parameters of cosmological perturbations we define the following functions
\begin{eqnarray}
\label{w1GD}
&& w_{1}= 1-\Delta_{ST}+2\frac{\Delta_{GB}}{H},\\
\label{w2GD}
&& w_{2}=2(1-\Delta_{ST})H-\dot{\Delta}_{ST}+6\Delta_{GB},\\
\label{w3GD}
&& w_{3}=-9(1-\Delta_{ST})H^{2}-3(1-\Delta_{ST})\dot{H}+\frac{3}{2}\ddot{\Delta}_{ST}-3\dot{\Delta}_{GB}
+\left(\frac{15}{2}\dot{\Delta}_{ST}-33\Delta_{GB}\right)H,\\
\label{w4GD}
&& w_{4}= 1-\Delta_{ST}-2\frac{\dot{H}}{H^{2}}\left(2\frac{\Delta_{GB}}{H}-\frac{\dot{\Delta}_{GB}}{\dot{H}}\right).
\end{eqnarray}

The velocities of scalar and tensor perturbations are
\begin{eqnarray}
\label{VS}
c_{S}^{2} & \equiv & \frac{3(2w_{1}^{2}w_{2}H-w_{2}^{2}w_{4}+4w_{1}\dot{w}_{1}w_{2}-2w_{1}^{2}\dot{w}_{2})}
{w_1(4w_{1}w_{3}+9w_{2}^{2})},\\
\label{VT}
c_{T}^{2} & \equiv & \frac{w_{4}}{w_{1}}.
\end{eqnarray}

If we consider a small deviation from GR, i.e.  $\Delta_{ST}\simeq0$ and $\Delta_{GB}\simeq0$ then we have
\begin{equation}
\label{w1GDE}
w_{1}= 1,~~~~~
w_{2}=2H,~~~~~
w_{3}=-9H^{2}-3\dot{H},~~~~~
w_{4}= 1,
\end{equation}
and from (\ref{VS})--(\ref{VT}) one can obtain $c_{S}\simeq1$ and $c_{T}\simeq1$, therefore the conditions for the crossing of the Hubble radius  $c_{S}k=aH$ and $c_{T}k=aH$, where $k$ is the wave number, are reduced to $k=aH$.

The power spectra which are given by the expressions
\begin{eqnarray}
\label{PSGEN}
{\mathcal P}_{S}=\frac{H^{2}}{8\pi^{2}Q_{S}c_{S}^{3}},~~~~~~
{\mathcal P}_{T}=\frac{H^{2}}{2\pi^{2}Q_{T}c_{T}^{3}},
\end{eqnarray}
with the following functions
\begin{eqnarray}
Q_{S}\equiv \frac{w_{1}(4w_{1}w_{3}+9w_{2}^{2})}{3w_{2}^{2}},~~~~~~
Q_{T} \equiv \frac{ w_{1}}{s},
\end{eqnarray}
for $\Delta_{ST}\simeq0$ and $\Delta_{GB}\simeq0$ can be written as
\begin{equation}
\label{PSGB}
{\mathcal P}_{S}=\frac{1}{2\epsilon}\left(\frac{H}{2\pi}\right)^{2},~~~~~~
{\mathcal P}_{T}=2s\left(\frac{H}{2\pi}\right)^{2}.
\end{equation}

The indices of scalar and tensor perturbations and tensor-to-scalar ratio for $\Delta_{ST}\simeq0$ and $\Delta_{GB}\simeq0$  are
\begin{equation}
\label{stilt}
n_{S}-1\equiv\frac{d\ln{\mathcal P}_{S}}{d\ln k}=
\frac{\dot{{\mathcal P}}_{S}}{H(1-\epsilon){\mathcal P}_{S}}=2\left(\frac{\delta-2\epsilon}{1-\epsilon}\right),
\end{equation}
\begin{equation}
\label{ttilt}
n_{T}\equiv\frac{d\ln{\mathcal P}_{T}}{d\ln k}=\frac{\dot{{\mathcal P}}_{T}}
{H(1-\epsilon){\mathcal P}_{T}}=-\frac{2\epsilon}{1-\epsilon},
\end{equation}
\begin{equation}
\label{RTGB}
r=\frac{{\mathcal P}_{T}}{{\mathcal P}_{S}}=4\frac{Q_{S}}{Q_{T}}\left(\frac{c_{S}}{c_{T}}\right)^{3}=
4s\epsilon,
\end{equation}
where $\delta=\epsilon-\frac{\dot{\epsilon}}{2H\epsilon}=-\frac{\ddot{H}}{2H\dot{H}}$ is the second slow-roll parameter.

In most works the parameters of cosmological perturbations are calculated for the value $s=4$ (for example, see~\cite{Linde:2005ht, Lyth:2009zz} and a lot of the other works on cosmological inflation). However, in some papers~\cite{Lukash:2006tr, Straumann:2005mz, Hwang:2005hb,Chervon:2008zz,Odintsov:2018zhw} one can find the other results which correspond to $s=1$. This difference can be easily explained by considering the normalization of the tensor of gravitational waves (tensor perturbations) which can be defined as~\cite{Lyth:2009zz,Straumann:2005mz}
\begin{equation}
\label{GWTensor}
\hat h_{ij}(\eta,\textbf{x}) = \int {\frac{d^3\textbf{k}}{(2\pi)^{3/2}}}
\sum_{\lambda=1,2}\left[h_k(\eta)\,e_{ij}(\textbf{k},\lambda)\,
\hat a_{\textbf{k},\lambda}\,e^{i{\textbf{k}}\cdot{\textbf{x}}} +
h^{\ast}_k(\eta)\,e^{\ast}_{ij}(\textbf{k},\lambda)\,
\hat a^{+}_{\textbf{k},\lambda}\,e^{-i{\textbf{k}}\cdot{\textbf{x}}}\right],
\end{equation}
by choosing the amplitude $h_k(\eta)=\sqrt{s}\times\overline{h}_k(\eta)$, where $\overline{h}_k(\eta)$ is the normalized
amplitude of gravitational waves.
In the expression (\ref{GWTensor}) $\textbf{k}$ is the wave vector, $\eta$ is the conformal time, $e_{ij}(\textbf{k},\lambda)$ are polarization tensors, $\hat a_{\textbf{k},\lambda}$ and $\hat a^{+}_{\textbf{k},\lambda}$ are the creation and annihilation operators, $\lambda=1,2$ characterizes two polarisations of gravitational waves.

Thus, two-point correlations for different amplitudes $\overline{h}_k(\eta)$ and $h_k(\eta)$ are related as
\begin{equation}
\sum_\lambda\langle0|h^*_{k,\lambda}h_{k',\lambda}|0\rangle=
s\times\sum_\lambda\langle0|\overline{h}^*_{k,\lambda}\overline{h}_{k',\lambda}|0\rangle ,
\end{equation}
and, therefore, from the connection between two-point correlation and power spectrum of tensor perturbations ~\cite{Lyth:2009zz,Straumann:2005mz}
\begin{equation}
\sum_\lambda\langle0|h^*_{k,\lambda}h_{k',\lambda}|0\rangle
\equiv \frac{(2\pi)^{3}}{4\pi k^{3}}{\mathcal P}_T(k)\delta^3(\textbf{k} - \textbf{k}'),
\end{equation}
one has $\mathcal{P}_T(k)=s\times\overline{{\mathcal P}}_T(k)$ which gives the different values of tensor-to-scalar ratio $r$ for $s=1$ and $s=4$.
To summarize the various results we will consider the parameters of cosmological perturbations in terms of the constant parameter $s=1,4$.

The observational constraints on the parameters of cosmological perturbations are~\cite{Aghanim:2018eyx}
\begin{eqnarray}
\label{PLANCK1}
&&{\mathcal P}_{S}=2.1\times10^{-9},~~~~~~~~~~~~~~~n_{s}=0.9663\pm 0.0041,\\
\label{PLANCK2}
&&r<0.1~~~\text{(Planck 2018)},~~~~~~r<0.065~~~\text{(Planck 2018/BICEP2/Keck-Array)}.
\end{eqnarray}

Thus, the parameters of cosmological perturbations for the case of GST gravity under the conditions (\ref{omegaSTGB})--(\ref{DEVGB}) coincide with ones in general relativity with high accuracy.
Therefore, the proposed approach leads to the generalisation of the inflationary models based on the Einstein gravity into the same ones in generalized scalar-tensor gravity.

\section{The examples of exact solvable GR-like cosmological models}

Now, we consider the the proposed approach on the example of the inflation with exponential potentials, namely
the power-law inflation and it's generalization as the exponential power-law inflation on the basis of the following Hubble parameter
\begin{equation}
\label{Hubble}
H(\phi)=\mu_{1}\exp(-\mu_{2}\phi)+\mu_{3},
\end{equation}
where $\mu_{1}$, $\mu_{2}$ and $\mu_{3}$ are some constants.

\subsection{The power-law inflation}

For the case of the power-law inflation we consider $\mu_{1}\neq0$, $\mu_{2}\neq0$ and $\mu_{3}=0$.

The exact solutions of the equations (\ref{E1GRLIKEF})--(\ref{E2GRLIKEF}) are
\begin{eqnarray}
\label{POTEPL}
&&V(\phi)=\mu^{2}_{1}(3-2\mu^{2}_{2})\exp(-2\mu_{2}\phi),\\
&&\phi(t)=\frac{1}{\mu_{2}}\ln\left[2\mu_{1}\mu^{2}_{2}t+c\right],\\
&&H(t)=\frac{\mu_{1}}{2\mu_{1}\mu^{2}_{2}t+c},\\
&&a(t)=a_{0}(2\mu_{1}\mu^{2}_{2}t+c)^{1/2\mu^{2}_{2}},
\end{eqnarray}
where $c$ is the constant of integration.

 Further, we calculate the exact explicit expression for coupling and kinetic functions from the equations (\ref{RCOR})--(\ref{Psi2})
 \begin{eqnarray}
 \label{fPL}
&&f(\phi)=-\frac{\beta_{ST}}{a^{2}_{0}}\exp\left(-\frac{\phi}{\mu_{2}}\right),\\
 \label{xiPL}
&&\xi(\phi)=-\frac{\alpha_{GB}}{a^{2}_{0}\mu^{3}_{1}(6\mu^{2}_{2}-5)}
\exp\left[\frac{(6\mu^{2}_{2}-5)}{2\mu_{2}}\phi\right]+const,~~\mu_{2}\neq\pm\sqrt{\frac{5}{6}}\\
 \label{omPL}
&&\omega(\phi)=1+\frac{3\exp\left(-\frac{5\phi}{2\mu^{2}}\right)}{2\mu_{1}\mu^{2}_{2}}
\left[\beta_{ST}\mu_{1}+2\alpha_{GB}\exp\left(-\frac{(3-2\mu^{2}_{2})}{2\mu_{2}}\phi\right)\right].
\end{eqnarray}

Therefore, we have generalized cosmological solutions which are reduced to GR ones for $\alpha_{GB}=0$ and $\beta_{ST}=0$.

\subsection{The exponential power-law inflation}

For the case of the exponential power-law inflation we consider $\mu_{1}\neq0$, $\mu_{2}\neq0$, $\mu_{3}\neq0$.

The exact solutions of the equations (\ref{E1GRLIKEF})--(\ref{E2GRLIKEF}) are
\begin{eqnarray}
\label{POTEPL2}
&&V(\phi)=\mu^{2}_{1}(3-2\mu^{2}_{2})\exp(-2\mu_{2}\phi)+6\mu_{1}\mu_{3}\exp(-\mu_{2}\phi)+3\mu^{2}_{3},\\
&&\phi(t)=\frac{1}{\mu_{2}}\ln\left[2\mu_{1}\mu^{2}_{2}t+c\right],\\
&&H(t)=\frac{\mu_{1}}{2\mu_{1}\mu^{2}_{2}t+c}+\mu_{3},\\
&&a(t)=a_{0}\exp(\mu_{3}t)(2\mu_{1}\mu^{2}_{2}t+c)^{1/2\mu^{2}_{2}}.
\end{eqnarray}

This model implies the exit from inflation, also for $t\rightarrow\infty$ the Hubble parameter $H=\mu_{3}$ which corresponds to the second accelerated exponential expansion of the universe.

 One can find the exact explicit expression for coupling and kinetic functions for $\mu_{2}=\sqrt{5/4}$ from (\ref{RCOR})--(\ref{Psi2})
\begin{equation}
\label{fPL2}
f(\phi)=-\frac{\beta_{ST}}{a^{2}_{0}}
\exp\left(\sqrt{-\frac{4}{5}}\phi-\frac{\mu_{3}}{\mu_{1}}e^{\sqrt{\frac{5}{4}}\phi}\right),
\end{equation}
\begin{equation}
\label{xiPL2}
\xi(\phi)=\frac{\alpha_{GB}}{5a^{2}_{0}\mu_{3}}
\left[\frac{\exp\left(-\frac{2\mu_{3}}{\mu_{1}}
e^{\sqrt{\frac{5}{4}}\phi}\right)}{\mu_{1}\left(\mu_{1}+\mu_{3}e^{\sqrt{\frac{5}{4}}\phi}\right)}-
\frac{2e^{2}}{\mu^{2}_{1}}{\rm Ei}\left(1,\frac{2\mu_{3}}{\mu_{1}}e^{\sqrt{\frac{5}{4}}\phi}+2\right)\right]+const,
\end{equation}
\begin{eqnarray}
\label{omPL2}
\nonumber
&&\omega(\phi)=1+\frac{6\left(\mu_{1}+\mu_{3}e^{\sqrt{\frac{5}{4}}\phi}\right)}{5\mu^{2}_{1}}
\exp\left(-\sqrt{\frac{4}{5}}\phi-\frac{4\mu_{3}}{5\mu_{1}}e^{\sqrt{\frac{5}{4}}}\right)\times\\
&&\left\{2\alpha_{GB}\exp\left[\left(\sqrt{\frac{4}{5}}-6\sqrt{5}\right)\phi-\frac{12\mu_{3}}{\mu_{1}}
e^{\sqrt{\frac{5}{4}}\phi}\right] +\beta_{ST}\left(\mu_{1}+\mu_{3}e^{\sqrt{\frac{5}{4}}\phi}\right)\right\}.
\end{eqnarray}
where ${\rm Ei}$ is the exponential integral function~\cite{Abramowitz}.

Also, the parameter $\mu_{2}=\sqrt{5/6}$ leads to the following exact solutions of the equations
 (\ref{RCOR})--(\ref{Psi2})
  \begin{equation}
\label{fPL1}
f(\phi)=-\frac{\beta_{ST}}{a^{2}_{0}}
\exp\left(-\sqrt{\frac{5}{6}}\phi-\frac{6\mu_{3}}{\mu_{1}}e^{\sqrt{\frac{5}{6}}\phi}\right),
\end{equation}
\begin{eqnarray}
\label{xiPL1}
\xi(\phi)=\frac{\alpha_{GB}}{5a^{2}_{0}\mu_{3}}
\left[\frac{\mu_{1}\exp\left(\frac{3\mu_{3}}{\mu_{1}}e^{\sqrt{\frac{5}{6}}\phi}\right)}
{\mu_{1}+3\mu_{3}e^{\sqrt{\frac{5}{6}}\phi}}+{\rm Ei}\left(1,\frac{3\mu_{3}}{\mu_{1}}e^{\sqrt{\frac{5}{6}}}\phi\right)
+\frac{2e^{3}}{\mu^{2}_{1}}{\rm Ei}\left(1,\frac{3\mu_{3}}{\mu_{1}}e^{\sqrt{\frac{5}{6}}}+3\right)\right]+const,
\end{eqnarray}
\begin{eqnarray}
\label{omPL1}
\nonumber
&&\omega(\phi)=1+\frac{9\left(\mu_{1}+\mu_{3}e^{\sqrt{\frac{5}{6}}\phi}\right)}{5\mu^{2}_{1}}
\exp\left(-\sqrt{\frac{5}{6}}\phi-\frac{6\mu_{3}}{\mu_{1}}e^{\sqrt{\frac{5}{6}}}\right)\times\\
&&\left\{2\alpha_{GB}\exp\left[\left(\sqrt{\frac{6}{5}}-9\sqrt{6}\right)\phi-\frac{54\mu_{3}}{\mu_{1}}
e^{\sqrt{\frac{5}{6}}\phi}\right] +\beta_{ST}\left(\mu_{1}+\mu_{3}e^{\sqrt{\frac{5}{6}}\phi}\right)\right\}.
\end{eqnarray}

 In this case, we have sufficiently complex expressions for coupling and kinetic functions to provide GR solutions
 for exponential power-law inflation in the case of GST gravity.

\subsection{The corresponding to observational constraints}

Further, we calculate the parameters of cosmological perturbations for this model for the arbitrary
value of $\mu_{2}$ and will consider  $\mu_{2}=\sqrt{5/4}$, $\mu_{2}=\sqrt{5/6}$ and $\mu_{3}=0$ as the partial cases.

Firstly, we obtain the expressions for slow-roll parameters
\begin{equation}
\label{SRPAR1}
\epsilon=-\frac{\dot{H}}{H^{2}}=
\frac{2\mu^{2}_{1}\mu^{2}_{2}}{(2\mu_{1}\mu^{2}_{2}\mu_{3}t+c\mu_{3}+\mu_{1})^{2}},
\end{equation}
\begin{equation}
\label{SRPAR2}
\delta=-\frac{\ddot{H}}{2H\dot{H}}=
\frac{2\mu_{1}\mu^{2}_{2}}{(2\mu_{1}\mu^{2}_{2}\mu_{3}t+c\mu_{3}+\mu_{1})}.
\end{equation}

For the case of the power-law inflation with $\mu_{3}=0$ from expressions (\ref{SRPAR1})--(\ref{SRPAR2}) we obtain
$\epsilon=\delta=2\mu^{2}_{2}$.

From the dependence of the e-folds number from cosmic time
\begin{equation}
N=\ln\left(\frac{a}{a_{0}}\right)=\frac{1}{2\mu^{2}_{2}}\ln(2\mu_{1}\mu^{2}_{2}t+c)
\end{equation}
we obtain the inverse relationship $t=t(N)$, namely
\begin{equation}
\label{tN1}
t(N)=\frac{1}{2\mu_{1}\mu^{2}_{2}}\left(e^{2\mu^{2}_{2}}-c\right).
\end{equation}

After substituting the dependence (\ref{tN1}) into the expression for power spectrum of the scalar perturbations on
the crossing on the Hubble radius (\ref{PSGB}) we have the condition
\begin{equation}
\label{cond1}
{\mathcal P}_{S}(N)=\frac{1}{2\epsilon}\left(\frac{H}{2\pi}\right)^{2}=
\frac{\mu^{2}_{1}}{16\pi^{2}\mu^{2}_{2}}\exp\left(-4\mu^{2}_{2}N\right)=2.1\times10^{-9}.
\end{equation}
The different choice of parameters  $\mu_{1}$ and $\mu_{2}\neq\sqrt{5/6}$ for $N=60$ can satisfy this condition.

From the expressions (\ref{stilt}) and (\ref{RTGB}) one has the following dependence $r=r(n_{S})$ for power-law inflation
\begin{equation}
\label{rnsPL1}
r=\frac{4s(n_{S}-1)}{n_{S}-3}.
\end{equation}

\begin{figure}[ht!]
\begin{minipage}[h]{0.455\linewidth}
\center{\includegraphics[width=1\linewidth]{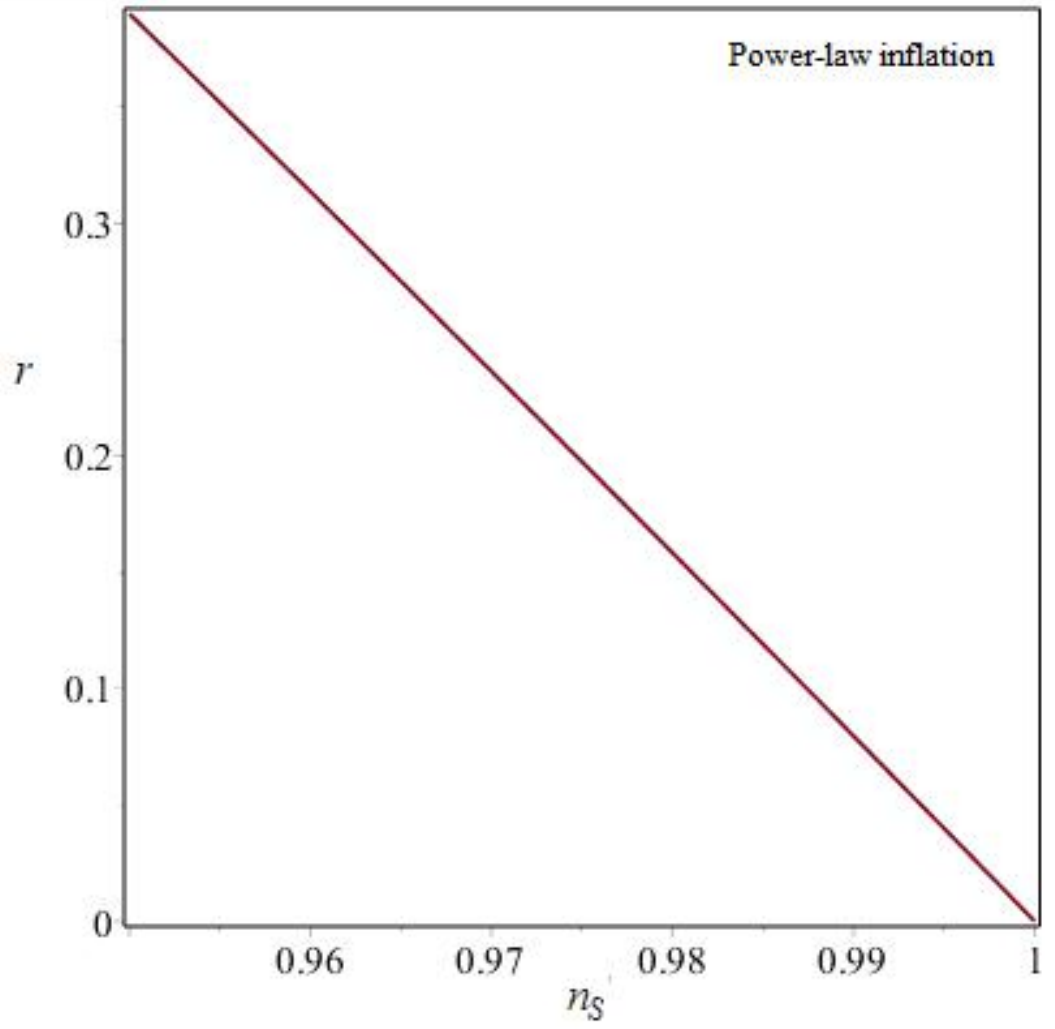}} \\
\end{minipage}
\hfill
\begin{minipage}[h]{0.47\linewidth}
\center{\includegraphics[width=1\linewidth]{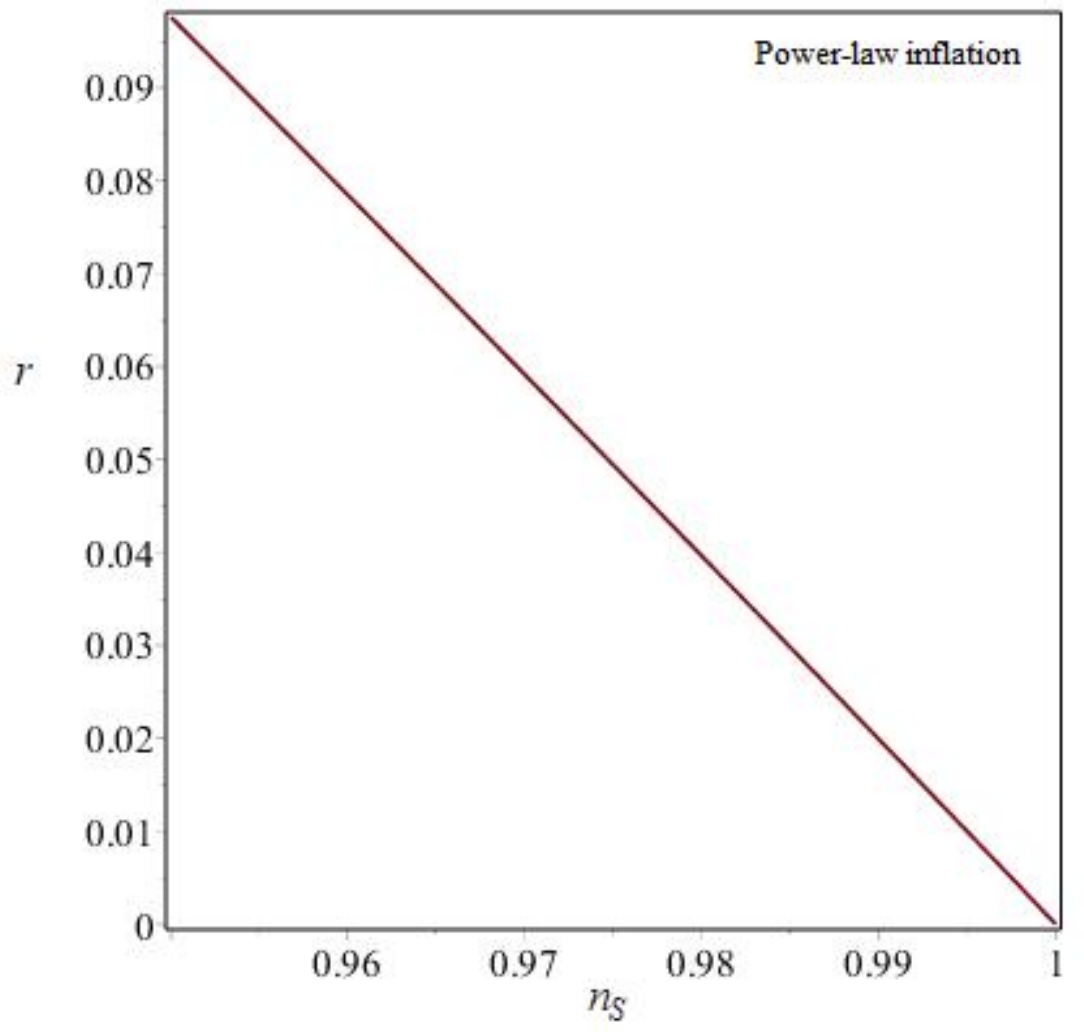}} \\
\end{minipage}
\vfill
\begin{minipage}[h]{0.47\linewidth}
\center{\includegraphics[width=1\linewidth]{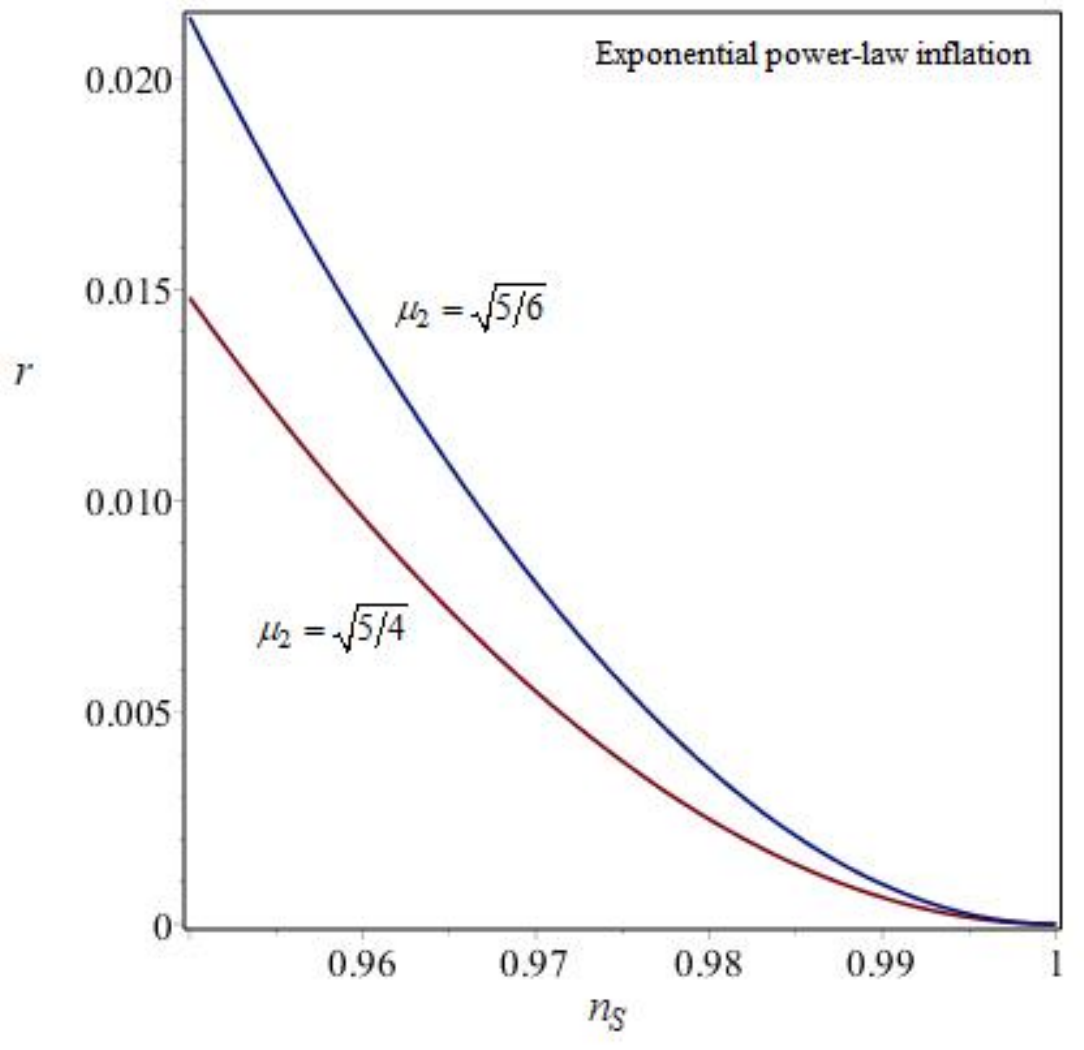}} s=4 \\
\end{minipage}
\hfill
\begin{minipage}[h]{0.47\linewidth}
\center{\includegraphics[width=1\linewidth]{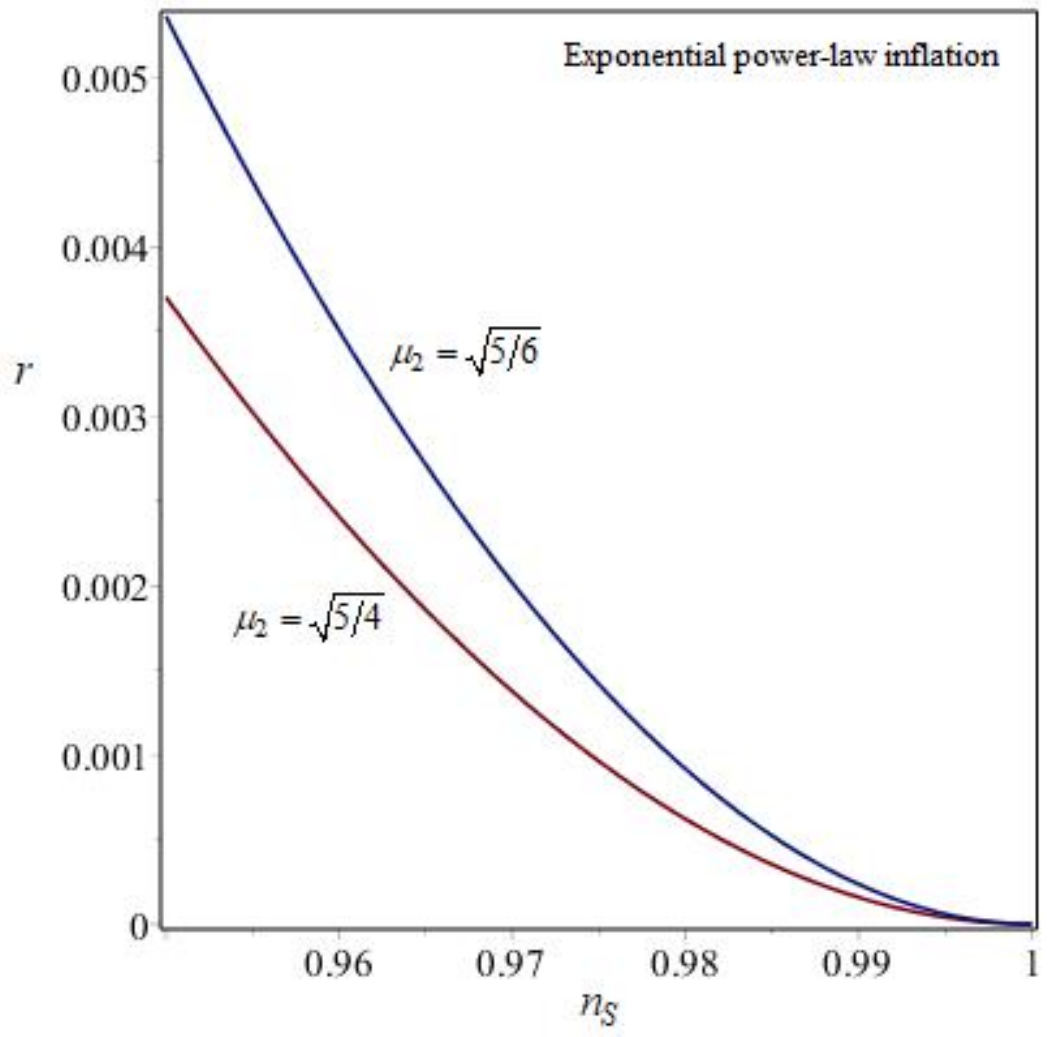}} s=1 \\
\end{minipage}
\caption{The dependences $r=r(n_{S})$ for exponential power-law inflation with $\mu_{2}=\sqrt{5/4}$, $\mu_{2}=\sqrt{5/6}$ and power-law inflation for both normalizations of the gravitational wave tensor $s=4$, $s=1$.}
\label{Fig1}
\end{figure}

For exponential power-law inflation with $\mu_{3}\neq0$ we have the condition
\begin{equation}
\epsilon=\frac{1}{2}\left(\frac{\delta}{\mu_{2}}\right)^{2}.
\end{equation}

Further, from the dependence of the e-folds number from cosmic time
\begin{equation}
N=\ln\left(\frac{a}{a_{0}}\right)=\frac{1}{2\mu^{2}_{2}}\ln(2\mu_{1}\mu^{2}_{2}t+c)+\mu_{3}t
\end{equation}
we obtain the inverse relationship $t=t(N)$, namely
\begin{equation}
\label{tN}
t(N)=-\frac{1}{2\mu_{1}\mu^{2}_{2}\mu_{3}}\left[c\mu_{3}-
\mu_{1}{\rm W}\left(\frac{\mu_{3}}{\mu_{1}}\exp(2\mu_{1}\mu^{2}_{2}N+c\mu_{3}{\mu_{1}})\right)\right],
\end{equation}
where ${\rm W}$ is denote the Lambert function~\cite{Abramowitz}.

After substituting the dependence (\ref{tN}) into the expression for power spectrum of the scalar perturbations on
the crossing on the Hubble radius (\ref{PSGB}) we have the condition
\begin{equation}
\label{cond}
{\mathcal P}_{S}(N)=\frac{1}{2\epsilon}\left(\frac{H}{2\pi}\right)^{2}=
\frac{\mu^{2}_{3}}{16\pi^{2}\mu^{2}_{2}}
\frac{\left[{\rm W}\left(\frac{\mu_{3}}{\mu_{1}}\exp(2N\mu_{1}\mu^{2}_{2}+c\mu_{3}{\mu_{1}})\right)+1\right]^{4}}
{\left[{\rm W}\left(\frac{\mu_{3}}{\mu_{1}}\exp(2N\mu_{1}\mu^{2}_{2}+c\mu_{3}{\mu_{1}})\right)\right]^{2}}=
2.1\times10^{-9}.
\end{equation}

As one can see, the different choice of four model's parameters  $c$, $\mu_{1}$, $\mu_{2}$, $\mu_{3}$ for $N=60$ can satisfy this condition.

Secondly, from (\ref{stilt}) and (\ref{RTGB}) we obtain the dependence
\begin{equation}
\label{rns}
r=\frac{16}{n_{S}-3}\left[n_{S}-1\mp\frac{\sqrt{2}\mu_{2}\left(-\sqrt{2}\mu_{2}
+\sqrt{2\mu^{2}_{2}+4n^{2}_{S}-16n_{S}+12}\,\right)}{2(n_{S}-3)}\right],
\end{equation}
where we will consider the solution with a negative sign.

From dependence $r=r(n_{S})$ on the Fig. \ref{Fig1} it can be concluded that the model of exponential power-law inflation corresponds restrictions (\ref{PLANCK1})--(\ref{PLANCK2}) for both normalizations $s=4$, $s=1$ and power-law inflation is not consistent with these restrictions for $s=4$. Consequently, the normalization of the gravitational wave tensor is critical for the verification of the power-law inflationary model on the spectral parameters of cosmological perturbations.

\section{The connection with Horndeski gravity}

Now, we consider the connection between the inflationary models which based on the action (\ref{actionh}) and
Horndeski gravity~\cite{Horndeski:1974wa} defined by the action
\begin{eqnarray}
&&S=\int d^4x\sqrt{-g}(L_2+L_3+L_4+L_5),\\
&&L_2=K(\phi,X),~~~L_3=-G_3(\phi,X)\Box\phi,\\
&&L_4=G_4(\phi,X)R+G_{4,X}\left[(\Box\phi)^2-(\nabla_\mu\nabla_\nu\phi)(\nabla^\mu\nabla^\nu\phi)\right],\\
&&L_5=G_5(\phi,X)G_{\mu\nu}\nabla^\mu\nabla^\nu\phi-\nonumber
\frac{1}{6}G_{5,X}[(\Box\phi)^3- \\
&&-3(\Box\phi)(\nabla_\mu\nabla_\nu\phi)(\nabla^\mu\nabla^\nu\phi)+
2(\nabla^\mu\nabla_\alpha\phi)(\nabla^\alpha\nabla_\beta\phi)(\nabla^\beta\nabla_\mu\phi)],
\end{eqnarray}
where $X=-\nabla_\mu\phi\nabla^\mu\phi/2$, $\Box\phi=\nabla_\mu\nabla^\mu\phi$,
$K$, $G_3$, $G_4$, $G_5$ is some functions of $\phi$ and $X$, $G_{j,X}(\phi,X)=\partial G_j(\phi,X)/\partial X$ with $j=4,5$.

For the following functions ~\cite{Mizuno:2010ag,DeFelice:2011uc}
\begin{eqnarray}
\label{HORN1}
&&K\left(\phi,X\right)=\omega(\phi) X-V(\phi)-4\xi''''_{\phi}X^2\left(3-\ln X\right),\\
&&G_3\left(\phi,X\right)=-2\xi'''_{\phi} X \left(7-3\ln X\right),\\
&&G_4\left(\phi,X\right)=\frac{1}{2}(1+f(\phi))-2\xi''_{\phi}X\left(2-\ln X\right),\\
\label{HORN4}
&&G_5\left(\phi,X\right)=2\xi'_{\phi}\ln X,
\end{eqnarray}
one can obtain the dynamic equations similar (\ref{beq2ah})--(\ref{beq4ah}) and, therefore, we have the same
cosmological models based on the Horndeski gravity.

Thus, to reconstruct the parameters of this type of gravity it is enough to calculate the derivatives of the function $\xi(\phi)$ and substitute the solutions of the equations (\ref{beq2ah})--(\ref{beq4ah}) in the expressions (\ref{HORN1})--(\ref{HORN4}).

For example, for the power-law inflation with $const=0$ in the expression (\ref{xiPL}) one has
\begin{equation}
\label{derivatives}
\frac{\partial^{(k)}\xi(\phi)}{\partial\phi^{k}}=\left[\frac{(6\mu^{2}_{2}-5)}{2\mu_{2}}\right]^{k}\xi(\phi),
~~~~\text{where}~~~~k=1,2,3,4.
\end{equation}

After substituting the solutions (\ref{POTEPL}), (\ref{fPL})--(\ref{omPL}) and (\ref{derivatives}) into the expressions
(\ref{HORN1})--(\ref{HORN4}) we obtain the parameters of the the Horndeski gravity corresponding to power-law inflation.
Also, one can reconstruct the type of the Horndeski gravity for exponential power-law inflation in the similar way.

\section{Evolution of the universe in GR-like cosmological models}

Now we will consider the scenario of the evolution of the universe, based on a special class of cosmological solutions for the generalized scalar-tensor theory of gravity.

At the beginning of the inflationary stage, the values of the deviation parameters $\Delta_{ST}$, $\Delta_{GB}$ and, respectively, the functions determining the non-minimal coupling of the scalar field and the curvature
$f(\phi)$ and $\xi(\phi)$ may be large enough to provide a connection with string and superstring theory~\cite{Zwiebach:1985uq}
which inspired the additional terms associated with modifications of the Einstein gravity.

During inflation, these deviations rapidly decrease according to the laws $\Delta_ {ST}\propto a ^{-2}(t)$ and $\Delta_{GB}\propto a^{-5}(t)$, thus, at the end of the inflationary stage, the part (\ref{FRF}) of the action (\ref{actionh}) which determines the type of gravity is
\begin{equation}
F(\phi,R)\equiv R+f(\phi)R+\xi(\phi)R^{2}_{GB}=R+\mathcal{O}_{1}\left(\Delta_{ST},\Delta_{GB}\right),
\end{equation}
where $\mathcal{O}_{1}\left(\Delta_{ST},\Delta_{GB}\right)\ll1$ is a small additive due to modifications of Einstein's gravity, which, when calculating the parameters of cosmological perturbations in a linear order perturbation theory, considered equal to zero.

The kinetic function at the end of the inflation is determined as follows
\begin{equation}
\omega(\phi)\equiv1+\frac{3}{\epsilon}\left(\Delta_{ST}+2\frac{\Delta_{GB}}{H}\right)=
1+\frac{3}{\epsilon}\mathcal{O}_{2}\left(\Delta_{ST},\Delta_{GB}\right),~~~
\mathcal{O}_{2}\left(\Delta_{ST},\Delta_{GB}\right)\ll1.
\end{equation}

Further, after completion of the inflation stage, the dynamics of the universe at the stages of the dominance of radiation and matter are determined by the Friedmann solutions $a(t)\propto t^{1/2}$ and $a(t)\propto t^{2/3}$, and the deviations with Einstein's gravity continues to decrease with the expansion of the universe.

At the present stage of the universe's evolution, i.e. for a larger number of e-folds, gravity in these models coincides with GR with even higher accuracy than in previous eras.
In the event that the stage of the second accelerated expansion of the Universe in the present epoch is described by
$\Lambda$CDM model with exponential expansion ($\dot{H}=0$) one has $\epsilon=0$ and $\omega\rightarrow\infty$, which corresponds to Einstein's gravity. If dark energy is determined by a quintessence field with a different corresponding dynamics  $\dot{H}\neq0$ ($\dot{H}\approx0$), then $\epsilon\neq0$ and the observational constraint $|\omega|>50000$~\cite{Bertotti:2003rm} defines small deviations from GR.
The second criterion limiting deviations from Einstein's gravity is the accuracy $ 10^{-15} $ in determining the propagation velocity of gravitational waves from the merger of neutron stars and black holes~\cite{Monitor:2017mdv}.

\section{Conclusion}

Investigations of modified gravity theories, including that with non-minimal coupling of a scalar field to Ricci and Gauss-Bonnet scalars, are connected with early inflationary epoch and with nowadays accelerated expansion of the universe (dark energy).

In the present paper, we studied the connection of exact solutions in GR cosmology with that dictated by modified 4D gravity with the self-interacting scalar field coupled to Ricci and Gauss-Bonnet scalars by minimal and non-minimal manners.

It was proved the existence  for each exact GR cosmology solution the same solution in the GSTG cosmology for the special choice of their parameters (\ref{RCOR})--(\ref{RCOR1}). Namely, we introduced the deviation parameters $\Delta_{ST}$ and $\Delta_{GB}$ (\ref{fST})--(\ref{xiGB}) as qualitative characteristic of difference between the models and found the connection of the functional parameters of GSTG with the scale factor of GR cosmology solutions.
By this approach a new class of exact cosmological solutions in models based on generalized scalar-tensor theories of gravity was obtained.
We also examined the connection of such models with the Hordeski gravity.

It was shown that the initially large deviation between Einstein's gravity and its considered modifications rapidly decreases with the expansion of the Universe. The parameters of cosmological perturbations in the proposed models coincide with ones in the case of general relativity with high accuracy.

We gave the algorithm of reconstruction for any exact solution obtained in GR~\cite{Chervon:2017kgn,Fomin:2017xlx} the same solution in GSTG cosmology
with the same parameters of cosmological perturbations.

As an example of this approach we obtained the exact cosmological solutions for power-law and exponential power-law inflation. 
The correspondence of these models to modern observational data on the spectral parameters of cosmological perturbations was  considered. The effect of the normalization of the gravitational wave tensor on the verification of cosmological models was also discussed.

\section{Acknowledgements}

This work has been partially supported by the RFBR grant 18-52-45016  IND a.
S.V.C. is grateful for support by the Program of Competitive Growth of Kazan Federal University.


\end{document}